\documentstyle[pre,aps,epsfig,multicol]{revtex}

\begin{document}
\title{Linear and nonlinear 
information flow \\ in spatially extended systems}
\author{Massimo Cencini $^{(a)}$ and Alessandro Torcini$^{(b),(c)}$}
\address{$(a)$ Max-Planck-Institut f\"ur Physik komplexer Systeme,
N\"othnitzer  Str. 38, D-01187 Dresden, Germany  } 
\address{$(b)$ Physics Department, Universit\'a ``La Sapienza'',
piazzale Aldo Moro 2, I-00185 Roma, Italy} 
\address{$(c)$ Istituto Nazionale di Fisica della Materia, UdR Firenze,
L.go E. Fermi, 3 - I-50125 Firenze, Italy}

\draft

\maketitle

\begin{abstract}
Infinitesimal and finite amplitude error propagation in 
spatially extended 
systems are numerically and theoretically investigated.
The information transport in these systems can be characterized
in terms of the propagation velocity of perturbations $V_p$.
A linear stability analysis is sufficient to capture all
the relevant aspects associated to propagation of 
infinitesimal disturbances. In particular, this 
analysis gives the propagation velocity $V_L$ of 
infinitesimal errors. If linear mechanisms prevail
on the nonlinear ones $V_p = V_L$. On the contrary, if 
nonlinear effects are predominant finite amplitude 
disturbances can eventually propagate faster than infinitesimal 
ones (i.e. $V_p > V_L$). The finite size Lyapunov exponent
can be successfully employed to discriminate the linear
or nonlinear origin of information flow.
A generalization of finite size Lyapunov exponent
to a comoving reference frame allows to state a
marginal stability criterion able to provide $V_p$
both in the linear and in the nonlinear case.
Strong analogies are found
between information spreading and propagation 
of fronts connecting steady states 
in reaction-diffusion systems. The analysis of
the common characteristics of these two phenomena 
leads to a better understanding of the role played
by linear and nonlinear mechanisms for the flow of 
information in spatially extended systems.
\end{abstract}
%
%

\begin{multicols}{2}
\section{Introduction}
\label{sec:1}
It is well recognized that chaotic dynamics generates a flow of
information in {\it bit space}: due to the sensitive dependence on initial
conditions one has an information flow from ``insignificant'' digits
towards ``significant'' ones \cite{Shaw}.  
In spatially distributed systems, due to the spatial coupling, one has an 
information flow both in {\it bit space} and in {\it real space}.
The flow in {\it bit space} is typically characterized in 
terms of the maximal growth $\lambda$ rate of infinitesimal disturbances
(i.e. of the maximal Lyapunov exponent), while the spatial information flow 
can be measured in terms of the maximal velocity of disturbance propagation
$V_p$ \cite{K86,BR87,grass89,PV94}. 

The evolution of a typical infinitesimal disturbance in low dimensional systems 
is fully determined once the maximal Lyapunov exponent is known. 
The situation is more complicated in spatio-temporal chaotic systems,
where infinitesimal perturbations can evolve both in time 
and in space. In this case a complete description of the dynamics
in the tangent space requires the introduction of other indicators,
e.g.: the comoving Lyapunov exponents \cite{DK87} and
the spatial and the specific Lyapunov spectra \cite{lepri}.

Nevertheless, the complete knowledge of  these Lyapunov spectra is not 
sufficient to fully characterize 
the irregular behaviors emerging in dynamical systems, this is
particularly true when the evolution of finite perturbations
is concerned.  Indeed, finite disturbances, which  are not confined 
in the tangent space, but are governed by the complete nonlinear 
dynamics, play a fundamental
role in the erratic behaviors observed in some high dimensional
system \cite{CK88,PLOK93,TP94,TGP95,CFVV99}. A rather 
intriguing phenomenon, termed {\it stable chaos}, 
has been reported in \cite{PLOK93}: the authors 
observed that even a linearly stable system 
(i.e. with $\lambda < 0$) can display an erratic behavior with $V_p>0$.

The first attempts to 
describe nonlinear perturbation evolution have been reported in
\cite{dressler92}. However, in these studies the analysis
was limited to the temporal growth rate associated with 
second order derivatives of one dimensional maps.
A considerable improvement 
along this direction has been recently achieved with the introduction of
the
finite size Lyapunov exponent (FSLE) \cite{ABCPV96}:
a generalization of the maximal Lyapunov exponent able to 
describe also finite amplitude perturbation evolution. 
In particular, the FSLE has been already demonstrated 
useful in investigating high dimensional systems \cite{CFVV99}.

The aim of this paper is to fully characterize the infinitesimal 
and finite amplitude perturbation evolution in spatio-temporal
chaotic systems. Coupled map lattices (CML's) \cite{cml} 
are employed to mimic spatially extended chaotic systems.
The FSLE is successfully applied to discriminate 
the linear or nonlinear origin of information propagation 
in CML's. Moreover, a generalization
of the FSLE to comoving reference frame (finite size comoving Lyapunov exponent)
allows to state
a marginal stability criterion able to predict $V_p$
in both cases: linear or nonlinear propagation.
A parallel with front propagation in reaction-diffusion \cite{KPP,wim2}
(non-chaotic) systems is worked out. The 
analogies between the two phenomena authorize to draw a 
correspondence between ``pulled'' (``pushed'') fronts 
and linear (nonlinear) information spreading.

The paper is organized as follows.
In Sect. II the FSLE is introduced
and applied to low dimensional systems (i.e. to single chaotic maps).
Sect. III is devoted to the description and 
comparison of linear and nonlinear 
disturbance propagation observed in different CML models.
The finite size comoving Lyapunov exponent is introduced in Sect. IV
and employed to introduce a generalized marginal stability criterion
for the determination of $V_p$. A discussion on information
propagation in non chaotic systems conclude Sect. IV.
The analogies between disturbance propagation in chaotic
systems and front propagation connecting steady states 
are analyzed in Sect. V.  The Appendix is devoted to the
estimation of finite time corrections for the computation of the 
FSLE in extended systems. Finally, some conclusive
remarks are reported in Sect. VI.

\section{Finite Size Lyapunov Exponent: \\
Low Dimensional Models}
\label{sec:2}

Let us introduce the FSLE by considering the
dynamical evolution of the state variable 
${\bf x}={\bf x}(t)$ ruled by
$$
{\dot {\bf x}}(t)={\bf f}[{\bf x}(t)] \, ,
$$
where ${\bf f}$ represents a chaotic flow in the phase space.
In order to evaluate the growth rate of non infinitesimal perturbations 
one can proceed as follows: a reference, ${\bf x}(t)$, and 
a perturbed, ${\bf x}^{'}(t)$, trajectories are considered.
The two orbits are initially placed at a distance 
$\delta(0)=\delta_{min}$, with $\delta_{min}\ll 1$,
assuming a certain norm  $\delta(t)=||{\bf x}^{'}(t)- {\bf x}(t)||$. 
In order to ensure that the perturbed orbit relaxes
on the attractor a first scratch integration is performed
for both the orbits until their distance has grown from $\delta_{min}$ 
to $\delta_0$ (where $1 >> \delta_0 >> \delta_{min})$. 
This transient ensures also the 
alignment of the perturbation along the direction of maximal expansion. 
Then the two trajectories are let to evolve and 
the growth of their distance $\delta(t)$ through different pre-assigned 
thresholds ($\delta_n=\delta_0 r^{n}$, with $n=0,\dots,N$ and typically
$1 < r \le 2$) is analyzed.

After the first threshold, $\delta_0$, is attained  
the times $\tau(\delta_n,r)$ required for 
$\delta(t)$ to grow from $\delta_n$ up to $\delta_{n+1}$
are registered.
When the largest threshold $\delta_N$ (which should be obviously chosen
smaller than the attractor size) is reached, the perturbed
trajectory is rescaled to the initial distance 
$\delta_{min}$ from the reference one.

By repeating the above procedure ${\cal N}$ times, 
for each threshold $\delta_n$, one obtains
a set of ``doubling'' times (this terminology is strictly
speaking correct only if $r=2$) 
$\{\tau_i(\delta_n,r)\}_{i=1,\dots,{\cal N}}$
and one can define the average of any observable
$A = A(t)$ on this set of doubling times as:
$$
 \langle A\rangle_e \equiv \frac{1}{\cal N} \sum_{i=1}^{\cal N} A_i\,,
$$
where $ A_i = A(\tau_i(\delta_n,r))$. 
The average $\langle \cdot \rangle_e$
does not coincide with an usual time average $\langle \cdot \rangle_t$
along a considered trajectory in the phase space, since
the doubling times typically depend on the 
considered point along the trajectory and on the threshold
$\delta_n$.  The two averages are linked 
(at least in the continuous case) 
via the following straightforward relationship~\cite{cencrep}
\begin{equation}
\langle A(t) \rangle_t = \frac{1}{T} \int_0^T dt A(t) = 
\frac{ \langle A \tau \rangle_e}{\langle \tau \rangle_e}
\label{eq:2.1}
\end{equation}
where  $T = \sum_{i=1,{\cal N}} \tau_i(\delta_n,r)$ and
$\langle \tau(\delta_n,r)\rangle_{e} = {T/\cal{N}}$.

A natural definition of the Finite Size Lyapunov Exponent
$\lambda(\delta_n)$ is the following~\cite{ABCPV96}:
\begin{equation}
\lambda(\delta_n) \equiv \left\langle{1 \over \tau(\delta_n,r)}\right\rangle_{t} \ln r \equiv
{1 \over \langle \tau(\delta_n,r) \rangle_{e}} \ln r \, .
\label{eq:2.2}
\end{equation}
The last equality stems from the relationship among the two
averages reported in (\ref{eq:2.1}).

In the limit of infinitesimal perturbation $\delta_n$
and of infinite $T$ (or $\cal N$) the FSLE converges to the
usual maximal Lyapunov exponent
\begin{equation}
\lim_{{\cal N} \to \infty}
\lim_{\delta_n \to 0} \lambda(\delta_n) = \lambda \, .
\label{eq:2.3}
\end{equation}
In practice, at small enough $\delta_n$,
$\lambda(\delta_n)$ displays a plateau
$\sim \lambda$.
Moreover, one can verify that $\lambda(\delta_n)$ is
independent of $r$, at least for not too large 
$r$~\cite{ABCPV96}.

In Eq.~(\ref{eq:2.2}) continuous time has been assumed, 
but discrete time is the most natural choice when 
experimental data sets (typically sampled at fixed intervals) 
are considered. In order to generalize the FSLE's definition
to the case of discrete time dynamical systems, let us consider 
the following map
$$ {\bf x}(t+1)={\bf F}({\bf x}(t)) \,,$$ 
where ${\bf x}$ is a continuous 
variable, and $t$ assumes integer values.
In this case $\tau(\delta_n,r)=\tau$ has simply to be interpreted as the minimum 
``integer'' time such that $\delta(\tau) \ge  \delta_{n+1}$, and, 
since now $\delta(\tau)/\delta_n$ is a fluctuating quantity, 
the following definition is obtained
\begin{equation}
\lambda(\delta_n) \equiv {1 \over \langle \tau(\delta_n,r) \rangle_{e}}
\left\langle \ln \left( \delta(\tau) \over \delta_n
 \right) 
\right\rangle_{e} \, .
\label{eq:2.5}
\end{equation}
A theoretical estimation of (\ref{eq:2.5}) is rarely possible,
and in most cases, one can only rely on a numerical
computation of $\lambda(\delta_n)$. However, 
in the following we will report two simple cases 
for which an approximate analytic expression for
the FSLE can be worked out.

Let us first consider the tent map 
$$ F(x)= 1- 2\left|x-\frac{1}{2}\right|$$
where $x \in [0:1]$. This is a one dimensional chaotic map,
since  $\lambda= \ln 2$ is positive.

\begin{figure}
\epsfxsize=8truecm
\epsfysize=6truecm
\epsfbox{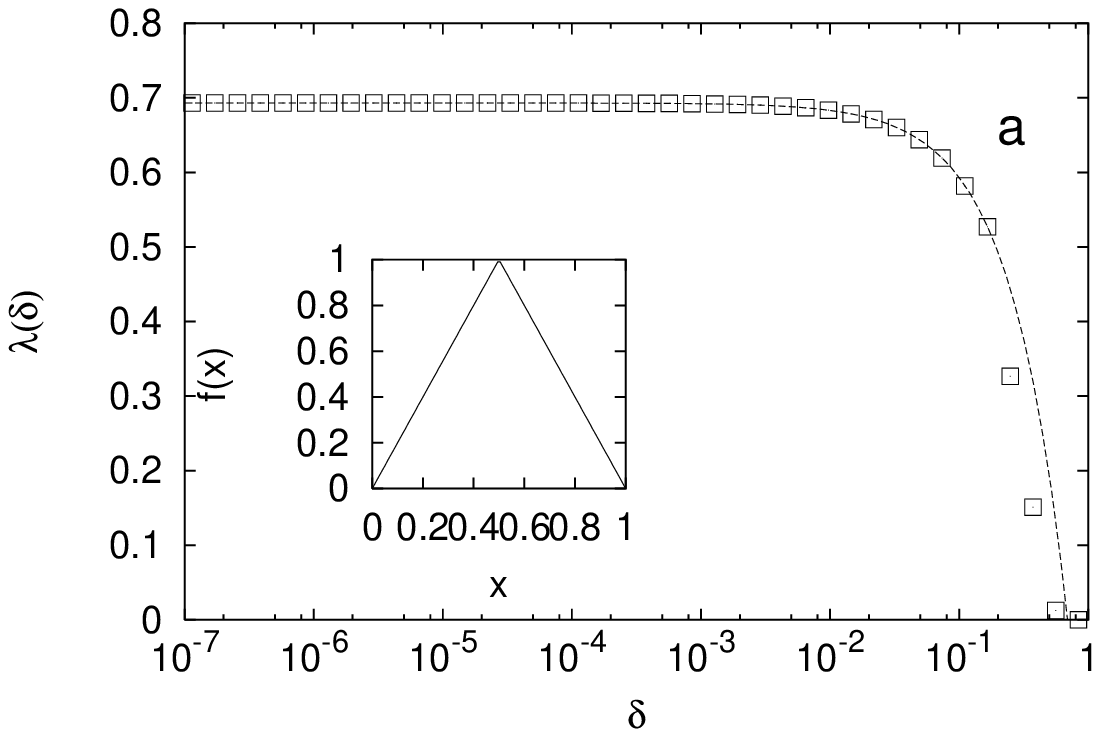}
\epsfxsize=8truecm
\epsfysize=6truecm
\epsfbox{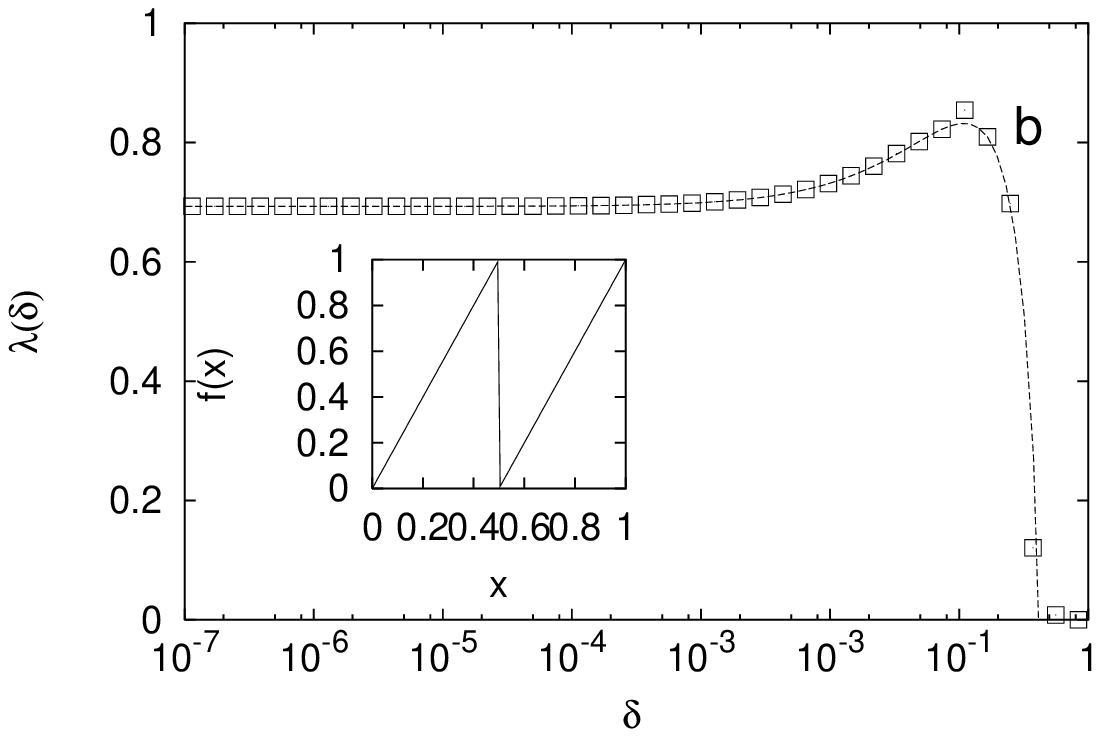}
\caption{$\lambda(\delta)$ versus $\delta$ for the tent map (a) and the 
shift map with $\beta=2$ (b). The continuous lines are the analytically 
computed FSLE and the boxes the numerically evaluated one. 
The two maps are displayed in the insets.  
}
\label{fig:maps}
\end{figure}

Due to the simplicity of this map, one can estimate the expression
(\ref{eq:2.5}) analytically obtaining the following
approximation 
\begin{equation}
\lambda(\delta) \simeq \ln 2 -\delta\,,
\label{eq:fsle_tent}
\end{equation}
valid for not too large $\delta$ values.
The maximal Lyapunov exponent is correctly recovered
in the limit $\delta \to 0$ and the above expression 
reproduces quite well the numerical estimate of the FSLE
(see Fig.~\ref{fig:maps} (a) ). An important point to stress
is that for this map the finite amplitude perturbations grow 
with the same rate or slower than the infinitesimal ones. 
The contraction of perturbations at large scales is
due to saturation effects related to the attractor
size. A similar dependence of the FSLE on the considered
scale is observed for the majority of the chaotic maps 
(logistic, cubic, etc.), as we have verified. 
One can wonder if there are systems for which, at variance with 
the behavior (\ref{eq:fsle_tent}), the finite size
corrections leads to an enhancement of the growth rate at 
large scales. As shown in \cite{TGP95}, the shift map
$F(x)= \beta x \; {\mbox {mod}} \; 1$ represents a good candidate. 
Also in this case it is possible to obtain
an analytical expression for the FSLE, when
$\delta < [1/(r+\beta)]$,
\begin{equation}
\lambda(\delta)=\frac{1}{1-\delta} \left[
(1-2\delta) \ln \beta+
\delta \ln \left({1-\beta\delta \over \delta}\right)
\right]\,,
\label{eq:lambda_shift}  
\end{equation}
which again correctly reproduces the numerical data
(see Fig.~\ref{fig:maps} (b) ) and in the limit
$\delta \to 0$ reduces to the corresponding maximal
Lyapunov exponent $\lambda=\ln \beta$.
As expected, finite amplitude disturbances can grow
faster than infinitesimal ones : 
\begin{equation}
\lambda(\delta) > \lambda(\delta \to 0)=\lambda \qquad {\mbox {for}} 
\:\: 0 < \delta \le \delta^{sat}\,,
\label{eq:nonlin}
\end{equation}
where $\delta^{sat}$ indicates the threshold at which saturation
effects set in. An even more interesting situation
is represented by the circle map 
$F(x)= \alpha + x \:\; {\mbox {mod}} \; 1$. This
map is marginally stable (i.e. $\lambda\!=\!\!0$),
but it is unstable at finite scales.
Indeed, the FSLE is given by 
$\lambda(\delta)\!\!=\!\!\delta/(1-\delta) \ln[(1-\delta)/\delta]$,
which is positive for $0 < \delta <1/2$. Therefore at small, but
finite, perturbations a positive growth rate is observed in spite 
of the (marginal) stability against infinitesimal perturbations. 
As a consequence, the circle map can exhibit behaviors that can
be hardly distinguished from chaos under the influence of
noise, since small perturbations may be occasionally driven 
into the nonlinear (unstable) regime and therefore amplified.
Of course, the role of noise can be played by coupling with
other maps, e.g. it has been found that coupled circle maps display 
behavior
resembling (for some aspects) that of a chaotic system \cite{TGP95}.
This phenomenon becomes even more striking in certain coupled stable maps where, 
even if the maximal Lyapunov exponent  is
negative \cite{PLOK93}, one can have a strong sensitivity to non infinitesimal
perturbations \cite{Kantz} (see  Sect.~\ref{sec:4.2} for a detailed discussion).

The two maps here examined for which (\ref{eq:nonlin}) holds
have a common characteristic: they are discontinuous.
However, in order to observe similar 
strong nonlinear effects, it is sufficient to consider a continuous map 
with high, but finite, first-derivative $|F^\prime|$ values 
\cite{TP94}. In this respect a simple example, that 
will be examined more in detail in Sect. IV, is represented 
by the map:
\begin{equation}
F(x)=\left \{ \begin{array}{cc}
b x & \qquad 0 \leq x < 1/b\\
1-c(1-q)(x-1/b) & \qquad 1/b \leq x < \frac{b+c}{bc}\\
q+d(x-\frac{b+c}{bc}) & \qquad \frac{b+c}{bc} \leq x \leq 1\,;
\end{array}
\right.
\label{eq:stable}
\end{equation}                                                 
with $b=2.7$, $d=0.1$, $q=0.07$ and $c=500$.
For $c \to \infty$ the map (\ref{eq:stable}) reduces to the one
studied in \cite{PLOK93}. For the map (\ref{eq:stable}) the FSLE 
dependence on $\delta$ is similar to that observed 
for the shift map.
\section{Information Spreading in Spatially Distributed Systems}
\label{sec:3}
In this section we will examine the mechanisms behind the
information flow in spatially distributed systems. 
In particular, the influence of
linear and nonlinear effects on information (error) spreading
will be analyzed. As a prototype of spatially 
distributed system Coupled Map Lattices (CML's) \cite{cml}
are considered :
\begin{eqnarray}
x_i(t+1)&=&F(\tilde{x}_i(t)) \nonumber \\
\tilde{x}_i(t)=(1-\varepsilon) x_i(t) &+& {\varepsilon \over 2} (x_{i-1}(t)+x_{i+1}(t))
\label{eq:cml}
\end{eqnarray}
where $t$ and $i$ are the discrete temporal and spatial indices, 
$L$ is the lattice size ($i=-L/2,\dots,L/2$), $x_i(t)$ the
state variable, and $\varepsilon \in [0:1]$ measures the
strength of the diffusive coupling. 
$F(x)$ is a nonlinear map of the interval 
ruling the local dynamics.

In order to understand how the information spreads
along the chain, let us consider two replicas of the same 
system, ${\bf x}(t)=\{x_i(t)\}$  and
${\bf x}^\prime(t)=\{x_i^\prime(t)\}$, 
that initially differ only in
a single site of the lattice (e.g. $i=0$) of a quantity $d_0$, i.e.
\begin{equation}
|x^\prime_i(0)-x_i(0)|=\Delta x_i(0)=d_0 \delta_{i,0}\,,
\label{eq:tzero}
\end{equation}
where $\delta_{i,0}$ is the Kronecker's delta. 
In a chaotic system the perturbation will typically
grow locally and spread along the chain.
These phenomena can be studied by considering the
difference field
\begin{equation}
\Delta x_i(t)= |x^\prime_i(t)-x_i(t)|=
|F(\tilde{x}^{'}_i(t-1))-F(\tilde{x}_i(t-1))|\,. 
\label{eq:dyn}
\end{equation}
It has to be stressed that the full nonlinear dynamics 
contributes to the evolution of $\Delta x_i(t)$.

\begin{figure}
\epsfxsize=8.truecm
\epsfysize=6truecm
\epsfbox{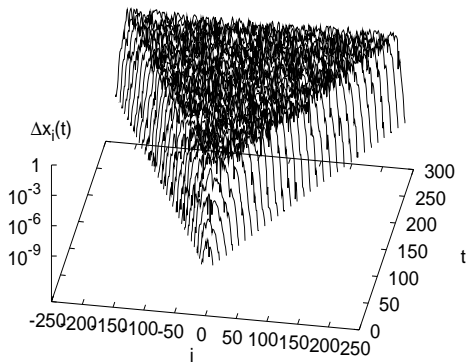}
\caption{Evolution of $\Delta x_i(t)$, 
for a chain of coupled tent map lattices
with a coupling $\varepsilon=2/3$.
The initial perturbation
is taken as in (\ref{eq:tzero}) with $d_0=10^{-8}$. 
}
\label{fig:torc.1}
\end{figure}

The spreading of this initially localized 
disturbance can be characterized in terms of
the velocity of information
propagation $V_P$ \cite{K86,grass89}.
As shown in Fig.~\ref{fig:torc.1}, $\Delta x_i(t)$ 
can grow only within a light-cone, determined by  $V_p$. 
For velocities higher than
$V_p$ the disturbance is instead damped. This individuates
a sort of predictability ``horizon'' in space-time: 
i.e. an interface separating 
the perturbed from the unperturbed region.
The velocity $V_p$ can be directly measured 
by detecting the leftmost, $i_l(t)$, and the rightmost,
$i_r(t)$, sites for which at time $t$ the perturbation 
$\Delta x_i(t)$ exceeds a preassigned
threshold. The definition of $V_p$ is the following
\begin{equation}
V_p=\lim_{t\to \infty}\lim_{L\to \infty} {i_r(t) -i_l(t)\over 2t}\,,
\label{eq:vel1}
\end{equation}
where  the limit $L\to \infty$ has to be taken first to avoid boundaries 
effects. The velocity (\ref{eq:vel1}) does not depend
on the chosen threshold values \cite{K86,grass89,TP94}.

Since the dynamics of the difference field 
(\ref{eq:dyn}) is not confined in the tangent space, 
non linearities can play a crucial role in the information propagation.
Indeed, we will see that the evolution of the disturbances strongly depend 
on the considered map $F(x)$ and in particular on the
shape of $\lambda(\delta)$. In the next subsection 
propagation in CML's with local chaotic
maps for which $\lambda(\delta)\leq \lambda \quad \forall \delta$
is discussed.  Local maps for which the condition 
(\ref{eq:nonlin}) holds will be the subject of 
Sect. \ref{sec:3.2}.
\subsection{Linear Mechanisms}
\label{sec:3.1}
Since in this subsection we consider 
CML's for which the local instabilities are 
essentially dominated by the behavior
of infinitesimal perturbations, 
most of the features can be understood
by limiting the analysis to the tangent space.

The evolution in tangent space is
obtained by linearizing 
Eq.~(\ref{eq:cml}), i.e.
\begin{eqnarray}
\delta x_{i}(t+1)&=&F^\prime(\tilde{x}_i(t)) [ \delta x_i(t)+
\nonumber\\
\frac{\varepsilon}{2}
(\delta x_{i+1}(t)&-&2 \delta x_{i}(t)+\delta x_{i-1}(t)) ]\,,
\label{eq:tang}
\end{eqnarray}
where $F^\prime$ is the first derivative of a one dimensional
chaotic map. Let us again consider as initial condition for
the evolution of (\ref{eq:tang}) a localized perturbation
as  (\ref{eq:tzero}) with $d_0$ infinitesimal.
The spatio-temporal dynamics of the 
tangent vector $\{ \delta x_i(t) \}$ is
determined by the interaction and competition of 
two different mechanisms present in Eq.~(\ref{eq:tang}):
the chaotic instability and the spatial diffusion.

As a first approximation, the effects of the 
two mechanisms can be treated as independent.
The chaotic instability leads to an average 
exponential growth of the infinitesimal disturbance:
$|\delta x_0(t)| \approx {d_0 \exp[\lambda t]}$.
On the other hand, the spatial diffusion, due to the
coupling, approximately leads to a 
spatial Gaussian spreading of the disturbance :
$|\delta x_i(t)| \approx 
|\delta x_0(t)| 
{/ \sqrt{4\pi Dt}} \exp( -{i^2 / 4Dt})$ 
where $D=\varepsilon/2$.
Combining these two effects one obtains
\begin{equation}
|\delta x_i(t)| \approx d_0 
{1 \over \sqrt{2\pi \varepsilon t}} \exp \left(\lambda t 
-{i^2 \over 2 \varepsilon t}\right)\,.
\label{eq:shape}
\end{equation}
Since the chaotic nature of the phenomenon
will typically  induces fluctuations, Eq.~(\ref{eq:shape}) can only describe 
the average shape of the disturbance. 
Moreover, Eq.~(\ref{eq:shape}) holds only 
when the perturbation is infinitesimal, since when
the disturbance reaches finite values a saturation mechanism 
(due to the nonlinearities) sets in preventing the divergence
of $|\delta x_i(t)|$. 

To verify the  validity of (\ref{eq:shape}), we studied the evolution
of localized perturbations of a homogeneous spatio-temporal 
chaotic state, in particular coupled logistic and tent maps
have been considered in the regime of ``fully developed turbulence''
\cite{K86}. Firstly, the system is randomly initialized and 
let to relax for a relatively long transient. 
At this stage two replicas of the same system are
generated and to one of the two a localized perturbation 
(like (\ref{eq:tzero})) is added. The evolution of the 
difference field (\ref{eq:dyn}) is then monitored at 
successive times.  In order to wash out the fluctuations,
the shape of the disturbance is obtained
averaging over many distinct realizations.
\begin{figure}
\epsfxsize=8.truecm
\epsfysize=6truecm
\epsfbox{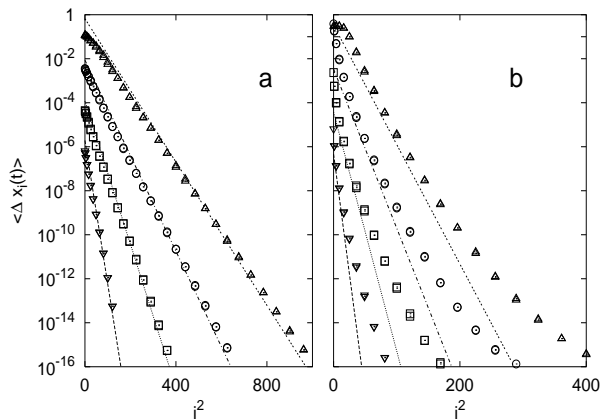}
\caption{Average evolution of perturbations for a CML of logistic maps
($f(x)=4x(1-x)$) for (a) $\varepsilon=1/3$ and (b) $\varepsilon=1/10$. 
$\langle \Delta x_i(t)\rangle$ is reported as a function of $i^2$ in a
lin-log scale at different 
times (from bottom to top $t=10,20,30,40$). Deviations from 
a straight line correspond to deviation from the Gaussian shape. 
$\langle \Delta x_i(t)\rangle$ is obtained as an average over $10^3$ realizations, 
for each one $\Delta x_i(0)$ has been chosen as (\ref{eq:tzero}) 
with $d_0=10^{-7}$.
For comparison the prediction (\ref{eq:shape})
is also reported (dashed lines).}
\label{fig:fronte}
\end{figure}
As one can see from Fig.~\ref{fig:fronte} Eq.~(\ref{eq:shape}) is fairly well 
verified for large enough coupling while it fails at small $\varepsilon$ \cite{nota}. 
These discrepancies are due to the finite spatial resolution
(that in CML's is always fixed to $1$), since
for small diffusivity constant the discretization of the Laplacian 
becomes inappropriate.
The expression (\ref{eq:shape}) for disturbance evolution
has been already proposed in Ref.~\cite{WB93}
for CML's in two dimensions. Deviations from
(\ref{eq:shape}) have been observed also in 
\cite{WB93}, but attributed to anomalous diffusive behaviors.
It has to be remarked that expression  (\ref{eq:shape})
is valid only at short times, since asymptotically ($t \to \infty$) 
the infinitesimal leading edge of the propagating front 
$|\delta x_i(t)|$ assumes
an exponential profile \cite{TGP95}.

For what concerns the propagation velocity,  
an estimation of $V_p$ can be obtained for infinitesimal
perturbations by the evaluation of the so-called maximal 
comoving Lyapunov exponents $\Lambda(v)$ \cite{DK87}. 
The time evolution of an intially localized (infinitesimal) 
disturbance (\ref{eq:tzero}) in a reference frame moving 
with velocity $v$ can be expressed as 
\begin{equation}
|\delta x_i(t)| \sim d_0 e^{\Lambda(v)t}\,,
\label{eq:growth}
\end{equation}
by following the perturbation along the world line $i=v t$
one can easily measures the corresponding
comoving Lyapunov exponent $\Lambda(v)$ (for more
 details see \cite{DK87,TP92}).
The information propagation velocity is the maximal
velocity for which a disturbance still propagates 
without being damped. Therefore it can be defined through
the following marginal stability criterion \cite{DK87}:
\begin{equation}
\Lambda(V_L) \equiv 0\,,
\label{eq:lvzero}
\end{equation}
where the velocity has been now indicated with $V_L$ in order
to stress that it has been obtained via 
a linear analysis. For the maps considered
in this section the identity $V_p = V_L$ is always fulfilled.

\begin{figure}
\epsfxsize=8.truecm
\epsfysize=6truecm
\epsfbox{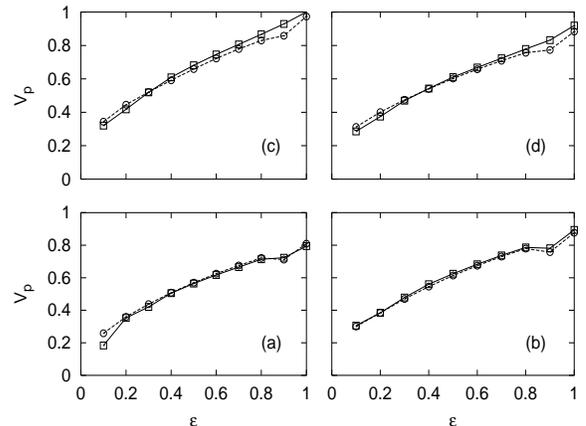}
\caption{Comparison between the directly measured propagation 
velocities $V_p = V_L$ (circles)
and the prediction (\ref{eq:velocity}) (boxes) for a CML of 
logistic maps with (a)
$a=3.9$ and (b) $a=4$, and of tent maps with (c) $a=2$ (d) $a=1.8$.
Lattices of $4\cdot 10^4$ maps has been used.}
\label{fig:vel}
\end{figure}
As shown in \cite{DK87}, in a closed system with symmetric coupling
$\Lambda=\Lambda(v)$ has typically a concave shape, 
with the maximum located at $v=0$ (in particular
$\lambda = \Lambda(v=0)$). An approximate expression 
can be obtained for $\Lambda(v)$, by substituting 
$i=vt$ in (\ref{eq:shape}) and by comparing
it with (\ref{eq:growth}) :
\begin{equation}
\Lambda(v)=\lambda-{v^2 / 2 \varepsilon} \quad .
\label{eq:comapr}
\end{equation} 
This parabolic expression for
$\Lambda(v)$ suffers of the same limits mentioned for 
the Gaussian approximation (\ref{eq:shape}) for the
disturbance evolution. 
Anyway, from (\ref{eq:comapr}) an analytical
prediction can be obtained for $V_L$ :
\begin{equation}
V_A=\sqrt{2\varepsilon \lambda}\,,
\label{eq:velocity}
\end{equation} 
which, as shown in  Fig.~\ref{fig:vel}, is indeed very good 
apart from some deviations for 
$\varepsilon \approx 0$ and $\varepsilon \approx 1$.
In Sect.~\ref{sec:5} we will rederive (\ref{eq:velocity}) 
by assuming that the chaotic perturbation 
behaves as a front connecting a stable and 
an unstable (metastable) fixed point
in a non-chaotic reaction diffusion system.

Let us briefly recall that another method (not suffering for
boundary problems) to determine the comoving Lyapunov exponent
has been introduced in~\cite{TP92}. The method relies on
the computation of {\it specific} Lyapunov exponents $\lambda(\mu)$ 
associated to an exponentially decaying perturbation (with spatial
decay rate $\mu$). In other words
one assumes that the spatio-temporal evolution of an infinitesimal
disturbance can be written as 
\begin{equation}
|\delta x_i(t)| \sim d_0 e^{\lambda(\mu)t-\mu i}\,.
\label{eq:specific}
\end{equation}
Since the asymptotic leading edge of the front separating perturbed 
from unperturbed part of the chain has an exponential shape,
the above assumption (\ref{eq:specific}) is appropriate to
study its evolution.

It is straightforward to show that the comoving Lyapunov
exponents are related to the specific ones via a Legendre
transform~\cite{TP92}, all the data concerning comoving exponents
reported in this paper have been obtained with such a method. Moreover,
a further results concerns the linear velocity $V_L$, it
can be shown \cite{TGP95} that
its value corresponds to the minimal propagation velocity
$V(\mu) = \lambda(\mu) / \mu$ associated with perturbations
of the form (\ref{eq:specific}), i.e.
\begin{equation}
V_L = \min_\mu \frac{\lambda(\mu)}{\mu} \equiv \frac{\lambda_L}
{\mu_L}
\label{eq:vmu}
\end{equation}
where $\mu_L$ and $\lambda_L=\lambda(\mu_L)$ represent the
spatial decay rate and the temporal growth rate of the leading
edge, respectively. The expression (\ref{eq:vmu}) for the linear
velocity is identical to the one derived for propagation of fronts
connecting a stable to an unstable steady state~\cite{wim1}. 

\subsection{Non Linear Mechanisms}
\label{sec:3.2}                   
In this section we investigate the case of coupled maps for which  
$\lambda(\delta) > \lambda(0)$ in some interval of $\delta$.
As noticed in Sect.~\ref{sec:2}, 
this behavior can be observed in chaotic (absolutely unstable) maps,
as well as in stable and marginally stable maps. Let us first
analyze chaotic maps, non-chaotic ones will be discussed
in Sect. \ref{sec:4.2}.

For these systems it is possible to have $V_p >  V_L$, 
this means that the disturbance can still propagate 
also in the velocity range $[V_L,V_p]$, 
even if the corresponding comoving Lyapunov exponents 
are negative. Therefore,
the linear marginal stability 
criterion (\ref{eq:lvzero}) does not hold anymore. 
We want to stress that the condition (\ref{eq:nonlin})
is necessary, but not sufficient to ensure that $V_p > V_L$,
since all the details of the coupled model play also an important role.

\begin{figure}
\epsfxsize=8.truecm
\epsfysize=6truecm
\epsfbox{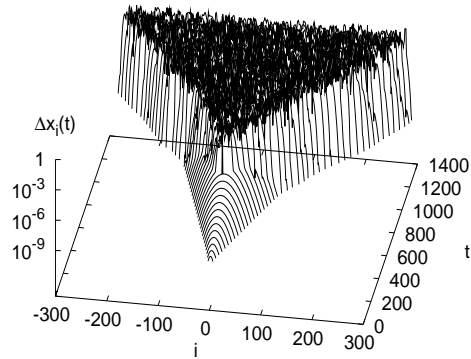}
\caption{The same of Fig.~\ref{fig:torc.1} for a lattice of coupled shift maps
with $\beta=1.03$,  $\varepsilon=1/3$.}
\label{fig:nonlinear}
\end{figure}

In Fig.~\ref{fig:nonlinear} the spatio-temporal evolution of an
initially localized disturbance of a chain of coupled shift
maps is reported.
As shown in \cite{isola}, when the coupling $\varepsilon \le 1/2$ the
maximal Lyapunov exponent for such model coincides with that
of the single map (namely, $\lambda=\ln(\beta)$) and if $\beta > 1$
the system is chaotic. Initially the perturbation, that is still
infinitesimal, spreads with the linear velocity $V_L$
above defined. At later time it
begins to propagate faster with a velocity $V_P > V_L$.
Comparing  Fig.~\ref{fig:nonlinear} with
Fig.~\ref{fig:torc.1}, one can see that the second stage of
the propagation sets in when the bulk of the perturbed region 
reaches sufficiently high values. As a matter of fact the initial 
stage of propagation disappears if we initialize the two replicas
with a disturbance of amplitude ${\cal O}(1)$. From these facts it is
evident that the origin of the information propagation characterized
by $V_p > V_L$ should be due to the strong nonlinear effects present
in this type of CML.

The behavior at long times can be understood by considering
the dependence of $\lambda(\delta)$ on the disturbance amplitude 
as shown in Fig.~\ref{fig:maps}b:
actually  the figure refers to the single map with $\beta=2$, but the 
shape of $\lambda(\delta)$ is qualitatively the same also for 
the coupled system and for other $\beta$-values.

\begin{figure}
\epsfxsize=8.truecm
\epsfysize=6truecm
\epsfbox{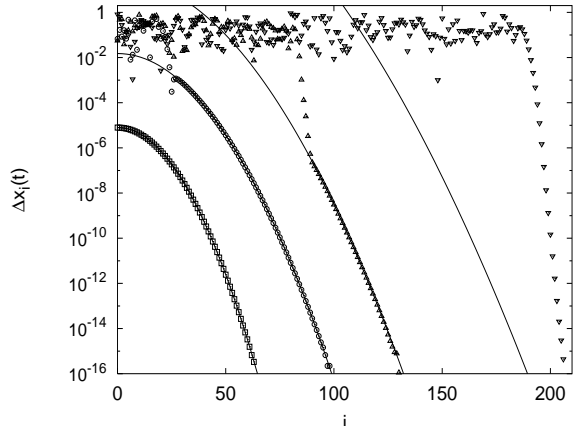}
\caption{Evolution of the perturbation $\Delta x_i(t)$ for a CML 
of  shift maps with
$\varepsilon=1/3$ and $\beta=1.04$ at four different times, $t=250,450,650,1000$.
The solid lines are the expected Gaussian  shape (\ref{eq:shape}).
The decay rate of the asymptotic exponential profile is 
$\mu_{NL} \sim 1.47$, noticeably greater than the linear 
value $\mu_L = 0.42$.}
\label{fig:invasion}
\end{figure}

Until the perturbation is infinitesimal $\lambda(\delta) \simeq \lambda$
and the linear analysis applies, when the disturbance becomes
bigger than a certain amplitude, $\delta^{NL}$, 
the growth will be faster, since
now $\lambda(\delta) > \lambda$. As it can be clearly seen in
Fig.~\ref{fig:invasion}, the perturbation is well reproduced by 
the linear approximation (\ref{eq:shape}) until 
the amplitude of the perturbation reaches a critical value
$\delta^{NL} \sim O(10^{-4})$ above which the nonlinear effects set in. 
At this stage
the nonlinear instabilities begin to push the front leading to
an increase of its velocity and deforming the profile of the perturbation.
This becomes exponential at much shorter times than in the linear
situation discussed in previous subsection. Moreover,
when the propagation is dominated by nonlinear mechanisms
the spatial decay rate $\mu_{NL}$ of the asymptotic leading edge
will be greater of the linear expected value $\mu_L$: this
result can be explained again invoking the analogy with
propagation of fronts connecting steady states \cite{TGP95}.
\begin{figure}
\epsfxsize=8.truecm
\epsfysize=6truecm
\epsfbox{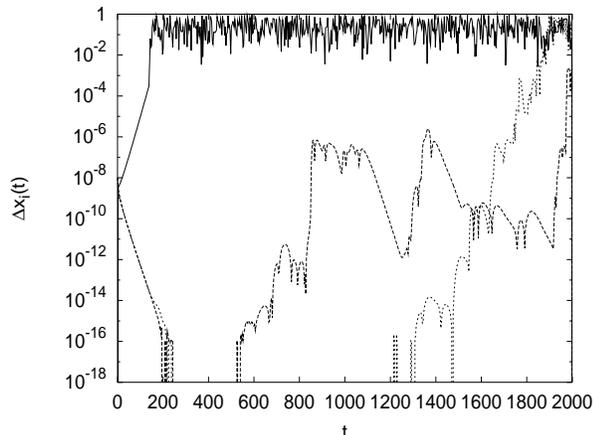}
\caption{Time evolution of $\Delta x_0(t)$ (solid line) and $\Delta
x_i(t)$ (dashed line), for $i=V_2 t$ with $V_2=1/3$. 
In this last case the data for two different chain configurations are reported.
The map used is the shift map with $\beta=1.1$, $\varepsilon=1/3$ and
$L=2\cdot 10^3$. For these parameters one has $V_L=0.250$ and $V_p=0.342$.
Note that between $t=200$ and $t=500$ for $\Delta x_i(t)$
the numerical precision is reached.}
\label{fig:bho}
\end{figure}
To better clarify the difference between the linear and non linear mechanisms
we show the behavior of $\Delta x_i(t)$ along  the world line $i=V t$ 
for two different velocities:  $V_1=0$ and $V_L<V_2<V_p$.
From Fig.~\ref{fig:bho}, one observes  for the zero-velocity situation
an exponential increase of the perturbation with rate $\lambda$
until it eventually saturates. For the case corresponding to
velocity $V_2$ an initial exponential decay with rate $\Lambda(V_2)$
is seen, followed at later times by a resurgence of the disturbance.
The successive evolution of the perturbation is no more exponential and 
exhibits strong fluctuations. 

These features suggest that in order to generalize the 
criterion (\ref{eq:lvzero}) to nonlinear driven information spreading
the growth of finite amplitude perturbations in a moving reference frame
should be analyzed.

\section{Finite Size Lyapunov Exponent: \\
Extended Systems}
\label{sec:4}
In this section we  introduce the 
Finite Size Comoving Lyapunov Exponent (FSCLE) that is
a generalization of the  FSLE to a moving reference frame.
First of all let us define the FSLE for an extended
system: in this case exactly the same definition
given in Sect.~\ref{sec:2} applies, apart from
some ambiguities in the choice of the
norm  to employ for measuring the distance
$\delta(t)$ of the perturbed, 
${\bf x}^\prime(t)=\{x_i^\prime(t)\}$, 
from the unperturbed replica, ${\bf x}(t)=\{x_i(t)\}$. 
A natural choice could be to perturb randomly
${\bf x}(t)$ and to look for the doubling times associated
to the evolution of the distance
\begin{equation}
{\tilde \Delta} (t) = \frac{1}{L} \sum_{i=-L/2}^{L/2} |x_i^\prime(t)-x_i(t)| 
\quad ;
\label{norm1}
\end{equation}
an alternative choice consists in
perturbing a single site of the chain, let's say $i=0$, 
at time $t=0$ and to evaluate the ``single site'' norm
\begin{equation}
\Delta x_0 (t) = |x_0^\prime(t)-x_0(t)| 
\quad .
\label{norm2}
\end{equation}
We have verified that the two norms (\ref{norm1}) and
(\ref{norm2}) give equivalent results for what concerns
the evaluation of $\lambda(\delta)$. In this paper
we will limit to consider the norm (\ref{norm2}).

In order to measure the FSCLE in a reference frame moving
with velocity $v$, we have simply measured
the difference (\ref{norm2}) along a world line $i=v t$;
i.e.
\begin{equation}
\Delta x_{i_c+[vt]} (t)  = |x_{i_c+[vt]}^\prime(t)-x_{i_c+[vt]}(t)| 
\quad .
\label{normv}
\end{equation}
where $[ \, \cdot \,  ]$ denotes the integer part and 
$i_c$ is introduced below.

The FSCLE is then estimated as in Sect.~\ref{sec:2}. Once 
a set of thresholds $\delta_n = r^n \delta_0$, with
$n=0,\dots,N$, is chosen and the perturbation is initialized 
as $\Delta x_i(0)=
\delta_{min} \, \delta_{i,0}$ with $\delta_{min}\ll \delta_0$. 
A preliminary transient evolution is performed in order to
allow to the perturbed orbit to relax on the attractor. 
At the end of this short transient the position $i_c$
where the perturbation reaches its maximum is detected,
this point is taken as the vertex of the light cone from
which all the considered world lines $i = v \, t$ depart.

The FSCLE is then defined as
\begin{equation}
\Lambda(\delta_n,v)={1 \over \langle \tau(\delta_n)\rangle_e} \left\langle 
\log\left({{\Delta} \over {\delta}_n}\right)
\right\rangle_e\,,
\end{equation}
where the dependence on the velocity $v$ derives from the employed 
norm (\ref{normv}). Note that we have used Eq.~(\ref{eq:2.5})
due to the discreteness of the temporal evolution.

In the limit of very small perturbation the FSCLE reduces to the comoving
Lyapunov exponent 
\begin{equation}
\lim_{\delta \to 0} \Lambda(\delta,v) =\Lambda(v)\,,
\label{eq:fslim}
\end{equation}
and, for $v=0$ one has the FSLE $\lambda(\delta)$. 
 
Actually there are finite time effects which prevents the limit 
(\ref{eq:fslim}) to be correctly attained.
This is related to the fact that the FSCLE's can be obtained 
only via finite time measurements. In the Appendix
we show how one can include such finite time corrections.

In the next subsections we will give evidences that 
the marginal {\it linear} stability criterion
(\ref{eq:lvzero}) can be generalized with the aid
of the FSCLE in the following way:
\begin{equation}
\max_{\delta} \{ \Lambda(\delta,v) \} =0\qquad {\mbox {for}} \, v \ge V_p\,,
\label{eq:crit}
\end{equation}
where $V_p$ can be either $V_L$, if the information
propagation is due to linear mechanisms, or greater
than $V_L$, when nonlinear mechanisms prevail on the
linear ones. For $v > V_p$ one has $\Lambda(\delta,v)=0$, 
since due to the definition of the
FSCLE negative growth rate appear to be $0$.
\subsection{Chaotic Systems}
\label{sec:4.1}
Let us first consider chaotic systems for which
$V_p \equiv V_L$, in this case 
$$\max_{\delta} \{ \Lambda(\delta,v) \} = \Lambda(v)$$
 and the generalized criterion (\ref{eq:crit}) reduces to the linear
one (\ref{eq:lvzero}). 

\begin{figure}
\epsfxsize=8.truecm
\epsfysize=6truecm
\epsfbox{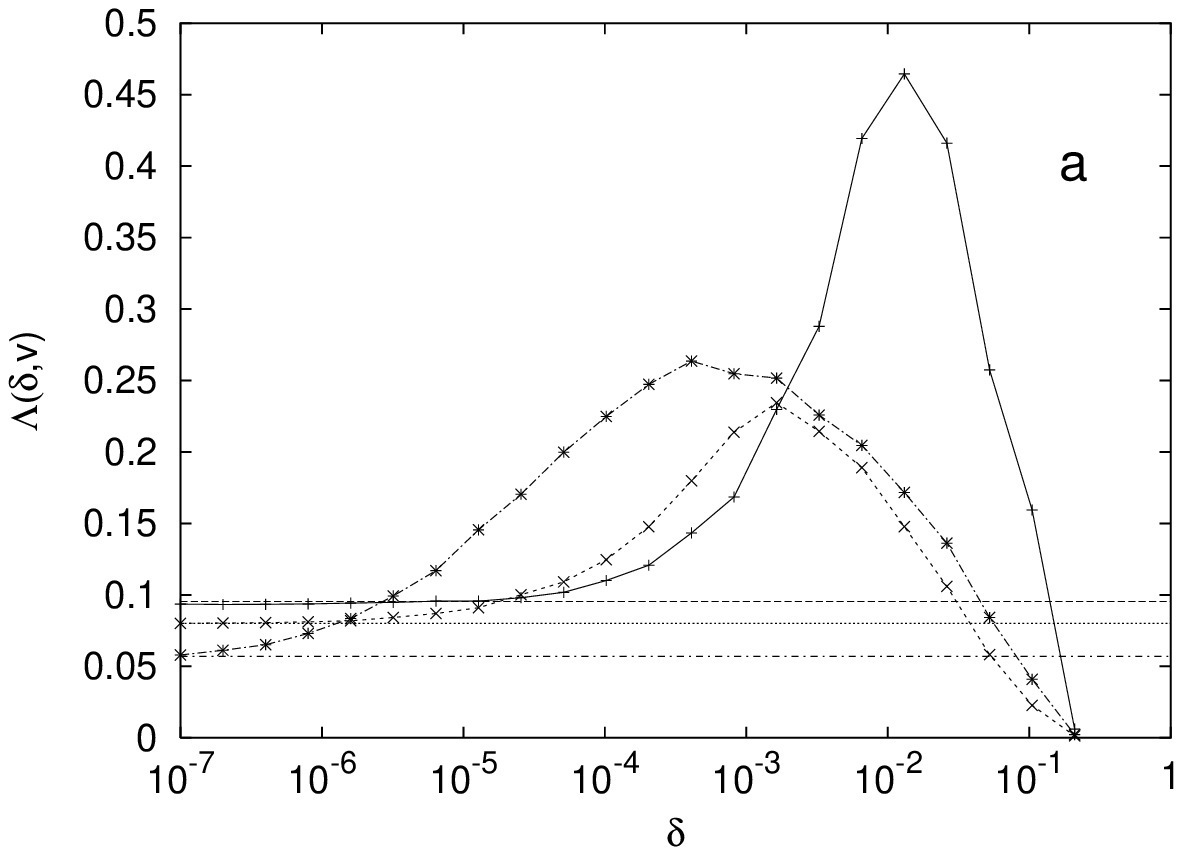}
\epsfxsize=8.truecm
\epsfysize=6truecm
\epsfbox{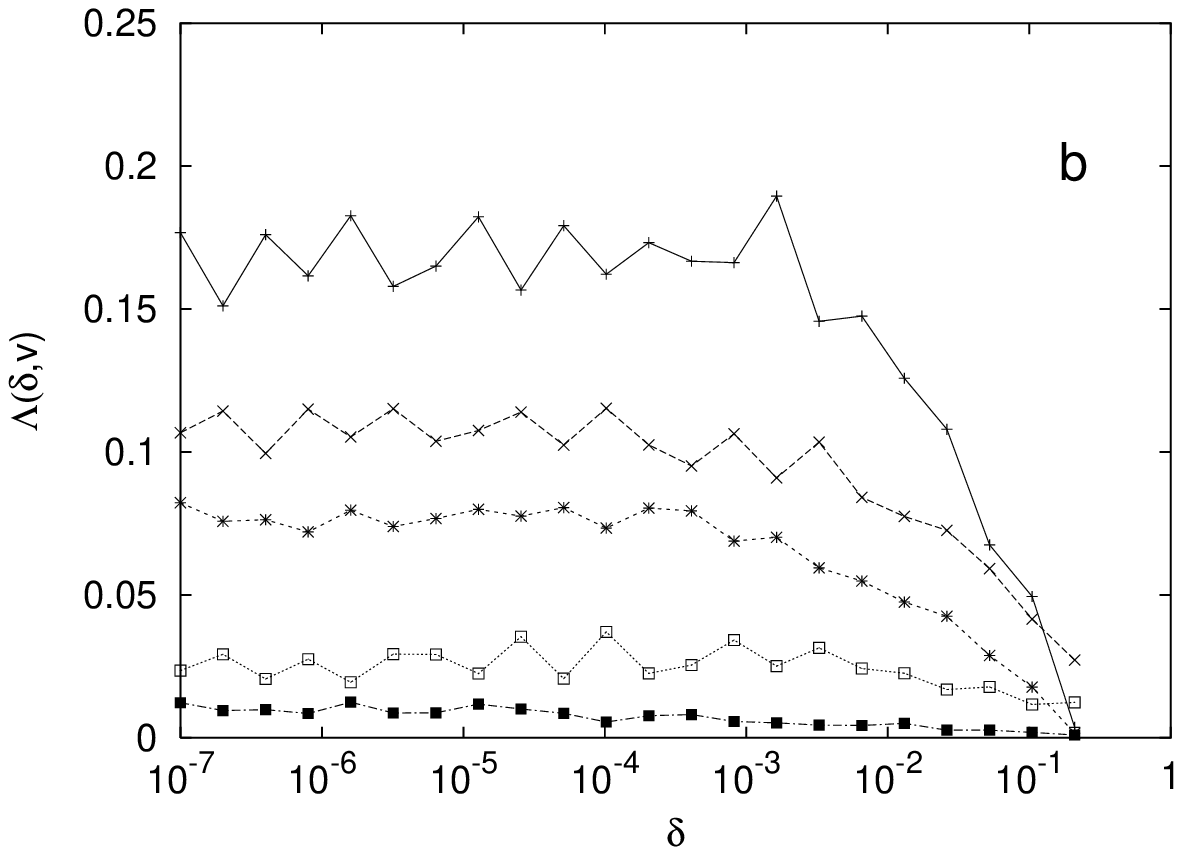}
\caption{$\Lambda(\delta,v)$ as a function $\delta$ 
for various velocities. The data
refer to the coupled shift maps with  $\beta=1.1$ and 
$\varepsilon=1/3$, with these parameters
$V_L \sim 0.250$ and $V_p \sim 0.342$.
In (a) results for velocities $v < V_L$ are reported,
namely (from top to bottom) $v=0$, $v=0.10$ and $v=0.16$. 
The straight lines indicate $\Lambda(v)$. 
In (b) velocities in the range $[V_L : V_p]$ are reported, from
the top $v=0.25$, $v=0.286$, $v=0.30$, $v=0.33$, $v=0.34$. 
Chains of lengths from $L=10^4$ up to $L=10^5$
have been employed and the statistics is 
over $2 \cdot 10^3$ doubling times.}
\label{fig:fscle1}
\end{figure}

As already stressed in Sect. \ref{sec:3.2}, shift coupled
maps represent a prototype of the class of chaotic models 
for which $V_p$ can be eventually bigger than $V_L$.
For this model the behavior of $\Lambda(\delta,v)$ for various 
velocities is reported in Fig.~\ref{fig:fscle1}.

For $v<V_L$, we observe that $\Lambda(\delta,v) \sim \Lambda(v)$ up to a 
certain value of the disturbance amplitude $\delta^{NL}$, above which
the FSCLE increases and exhibits a clear peak at some higher $\delta$-value.
From Fig.~\ref{fig:fscle1} (a) it is clear that
$\delta^{NL}$ (which denotes the set in of the regime dominated
by nonlinear mechanisms) decreases for increasing velocities and finally
vanishes at $v = V_L$. This behavior is reasonable since, as shown in 
Fig.~\ref{fig:nonlinear}, initially the disturbance evolves
along the chain following the linear mechanism 
characterized by $\Lambda(v)$, but as soon as in the central site
$\delta > \delta^{NL}$ the nonlinear mechanism begins to be active. 
Thus a nonlinear front is excited and this invades the linear region
propagating with a velocity higher than $V_L$, therefore the higher is $v$ 
the smaller is the scale at which nonlinear effects are observed.

\begin{figure}
\epsfxsize=8.truecm
\epsfysize=6truecm
\epsfbox{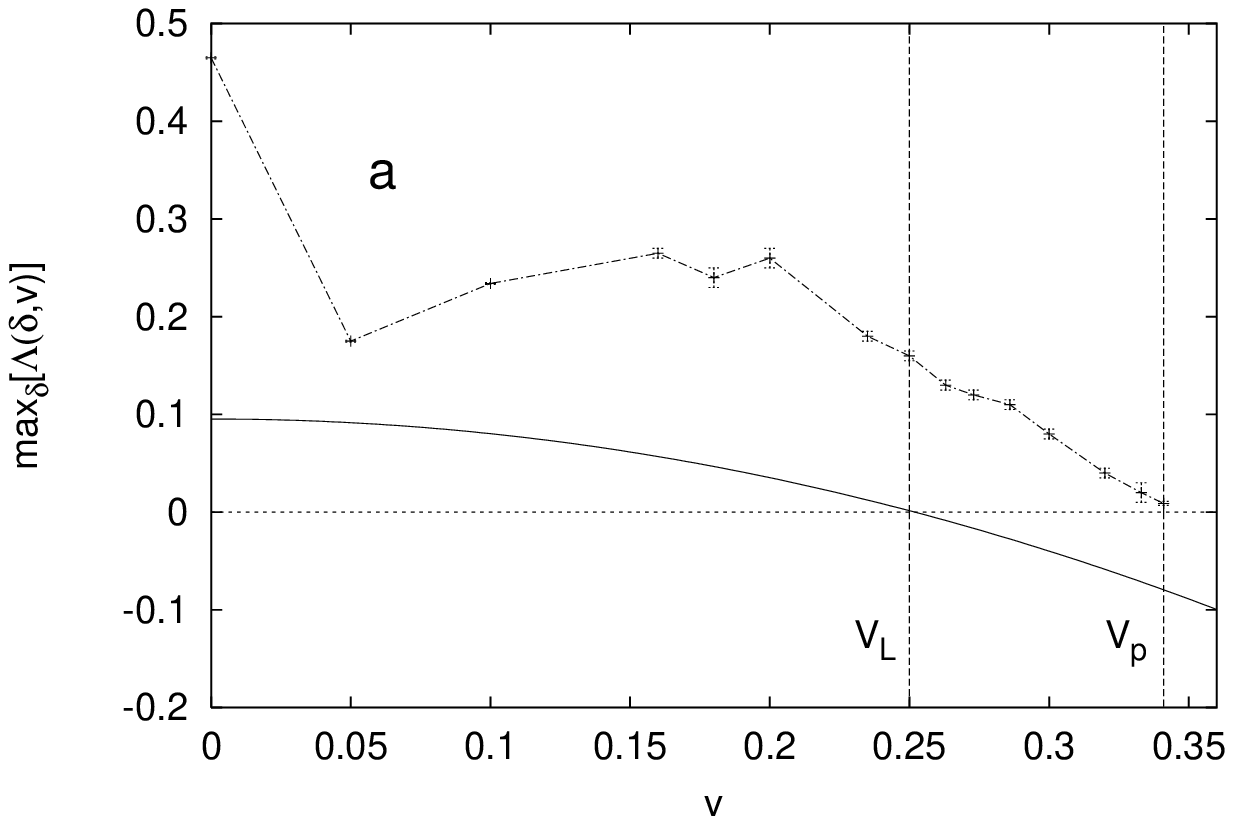}
\epsfxsize=8.truecm
\epsfysize=6truecm
\epsfbox{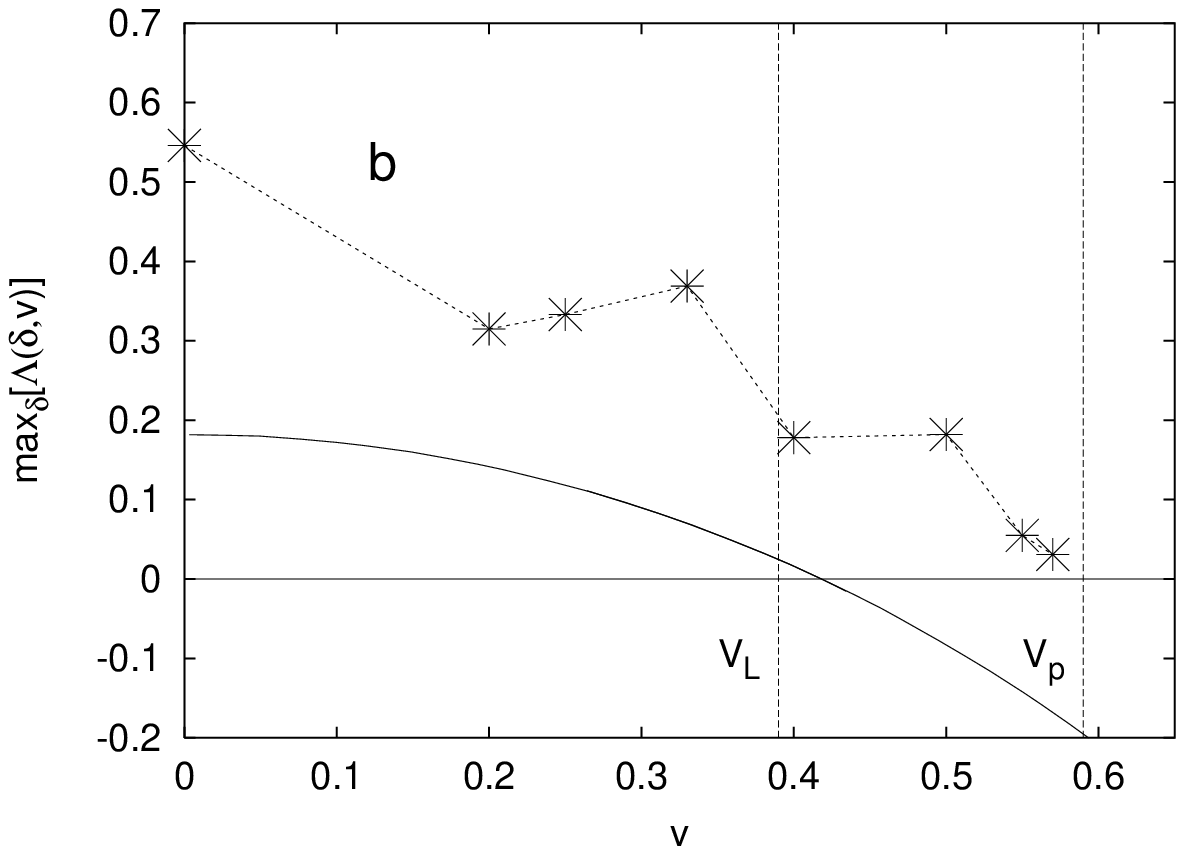}
\caption{$\max_{\delta}\Lambda(\delta,v)$ 
(dashed line with points) versus $v$ 
compared with $\Lambda(v)$ (continuous line).
(a) Results for coupled shift maps with the same parameters as Fig.~\ref{fig:fscle1},  
the vertical lines indicates  $V_L \approx 0.250$ and the 
measured (\ref{eq:vel1}) propagation velocity $V_p \approx 0.342$. 
(b) Data for the coupled maps (\ref{eq:stable}) with parameters 
$b=2.7$, $d=0.1$, $q=0.07$ 
and $c=500$ and with coupling $\varepsilon=2/3$,
the vertical lines indicates $V_L
\approx 0.39$ and $V_p \approx 0.59$. 
In this second case chains of length $L=2\cdot 10^4$ have been considered
and the statistics is over $2 \cdot 10^4$ doubling times.                            
The reported values for $\max_{\delta}\Lambda(\delta,v)$ 
refer to an average over five values of $\delta$ around the 
peak position of $\Lambda(\delta,v)$.
}
\label{fig:vlvnl}  
\end{figure}

In the interval $V_L \leq v \leq V_p$ (see Fig.~\ref{fig:fscle1}(b)), 
$\Lambda(\delta,v)$ is still 
positive and its maximal value decreases for increasing $v$. 
The FSCLE vanishes for $v=V_p$ as Fig.~\ref{fig:vlvnl} (a) shows. 
In this velocity range $\Lambda(v)$ is always negative,
therefore the instabilities observed in a reference frame
moving with velocity higher than $V_L$ have a fully nonlinear
origin. The residual fluctuations present in $\Lambda(\delta,v)$ are 
essentially due to the limited statistics.
Indeed the nonlinear growth as shown in
Fig.~\ref{fig:bho} is extremely fluctuating.

In Figure~\ref{fig:vlvnl}(a) the dependence of 
$\max_{\delta}\{\Lambda(\delta,v)\}$ on $v$ for coupled shift maps is 
reported, as expected it vanishes exactly for $v=V_p$.  
For $v < V_L$ a non-monotonous behavior of
$\max_{\delta}\{\Lambda(\delta,v)\}$ is observable,
this is probably due to the complex interplay of linear and nonlinear
effects. For $v > V_L$, 
the behavior is smoother and a monotonous decrease is observed.
As discussed in Sect.~\ref{sec:2}, the discontinuity present
in the shift map is not necessary in order to observe
nonlinear mechanisms prevailing on linear ones.

As an example of continuous map exhibiting an information
propagation velocity $V_p > V_L$
the map (\ref{eq:stable}) is considered. 
This type of CML has been already studied in \cite{TP94}:
it has been observed that in a certain parameter range
$V_p$ can be finite even if the map is non chaotic. 
Moreover, 
also in the chaotic regime there is a window of parameters
where $V_p > V_L$, with our parameters choice $\lambda \approx 0.182>0$.


Also in the present case we observe an overall behavior
of the FSCLE resembling that of the coupled shift maps.
The main point that we want to remark is that in the limit
$v \to V_p$  $\Lambda(\delta,v) \to 0$,
as shown in Fig.~\ref{fig:vlvnl} (b). 
At variance with the shift map (that is an everywhere
expanding map) the considered map shows contracting 
and expanding intervals. Therefore the disturbances
during their spatio-temporal evolution can be
alternatively expanded or contracted. This leads
to strong fluctuations in the FSCLE-values, that are 
difficult to remove. 

As a matter of fact we observed
that even for velocities slightly larger than $V_p$
$\Lambda(\delta,v)$ can be non zero. But the statistics
of these anomalous fluctuations is extremely low,
referring to the parameters reported in Fig.~\ref{fig:vlvnl} (b)
for $v=0.6 > V_p = 0.59$ we observed an expansion (instead
of the expected contraction) of disturbances in the 
$0.5 \%$ of the studied
cases. For higher velocities $\Lambda(\delta,v)$ is
zero for any considered $\delta$.

\subsection{Non Chaotic Systems}
\label{sec:4.2}
On the basis of the linear analysis discussed in Sect.~\ref{sec:3.1}
the propagation of disturbances should be present only
in chaotic systems. 
For non chaotic ones $\lambda \leq 0$ and from Eq.~(\ref{eq:lvzero}) 
one trivially obtains $V_L=0$. 
On the other hand, in the class of systems for which 
$\lambda(\delta)\geq \lambda$ 
there are also stable and marginally stable systems. Therefore,
a propagation due solely to non linear terms can
still be present \cite{PLOK93,TP94,TGP95}. 
In this section we want to discuss 
the possible employ of the FSLE  to characterize these maps.

Before entering into the description of the FSLE computation in such systems,
it is of interest to recall an important phenomenon which appears in 
stable systems: the so-called ``stable chaos'' \cite{PLOK93}. 
Stable systems asymptotically  evolve towards 
trivial attractors (i.e. fixed points, periodic or quasi-periodic orbits).
However, in spatially extended systems it may happen that the time needed 
to reach the asymptotic state is very long: it has been found that in certain 
stable CML the transient time diverges exponentially with the number of 
elements of the chain \cite{CK88}. 
Moreover, this transient is characterized by a quasi-stationary
behavior allowing for the investigation of the properties of the 
model with statistical consistency. In Ref. \cite{PLOK93}
it has been shown that a chain of coupled maps of the type (\ref{eq:stable}),
considered in their discontinuous limit (i.e. for $c \to \infty$),
is non chaotic but still exhibits erratic behaviors. This
is associated to a non-zero information spreading within
the system.

As far as the computation of the FSLE for these systems is concerned,
some remarks are worth to be done. The definition of the
FSLE in term of error doubling times cannot be used 
in a straightforward manner to determine negative expansion rates.
Another important point is that, at variance with the case 
of chaotic maps, finite perturbations should be now considered
in order to observe an expansion.

These two points impede a straightforward implementation of our method to
study these systems. Indeed, if one starts with too small perturbations  
the propagation does not manifest, while if one initializes the system with a
finite perturbation 
$\lambda(\delta)$ cannot be estimated with the required accuracy. 
Indeed, the only way to have an independence of $\lambda(\delta)$ from 
the initial conditions is to initialize the system with infinitesimal
perturbations and then to follow them until they become finite due to
the dynamics of the system (see Ref.~\cite{ABCPV96} for a detailed discussion).

The coupled (\ref{eq:stable}) maps for $c\to \infty$ have been recently
analyzed by Letz \& Kantz \cite{Kantz} in term of 
an indicator similar to the FSLE (i.e. able to quantify 
the growth rate of non infinitesimal perturbations). This
indicator turns out to be negative for infinitesimal 
perturbations and becomes positive for finite perturbations.
This means that a finite perturbation of sufficient amplitude 
can propagate along the system due to nonlinear effects.
This confirms what was previously observed
in Ref.~\cite{TGP95} for marginally stable systems.

\begin{figure}
\epsfxsize=8.truecm
\epsfysize=6truecm
\epsfbox{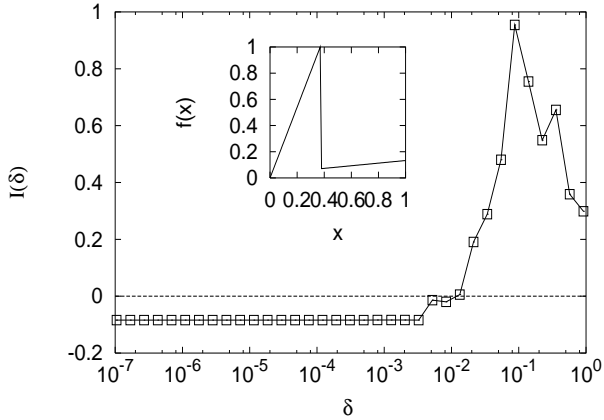}
\caption{$I(\delta)$ as a
function of the amplitude perturbation in
a lin-log scale for the discontinuous coupled 
maps (\ref{eq:stable}), in the limit $c \to \infty$, with
$\varepsilon=1/3$ and $L=35$. The map is reported in the inset.
The negative maximal Lyapunov $\lambda = -0.105$ is recovered
at small scales. For the computation details see the text.
In the inset the single map is shown.}
\label{fig:stable}                                      
\end{figure}
In Fig.~\ref{fig:stable} we show the behavior of 
a quantity $I(\delta_n)$ similar to $\lambda(\delta_n)$ which 
has been obtained as follows.
We considered two trajectories at an initial distance $\delta_n$,
after one time step evolution the distance $\delta$ 
between the trajectory is measured. Then one of the
two trajectory is rescaled at a distance 
$\delta_n$ from the other, keeping the direction of the perturbation unchanged,
and the procedure is repeated several times and for several
values of $\delta_n$.
Then we averaged $\ln(\delta/\delta_n)$ over many different initial conditions
obtaining $I(\delta_n)$. For $\delta_n \to 0$, this is nothing 
but  the usual algorithm for computing the maximal Lyapunov exponent~\cite{benettin}.

As discussed in Ref.~\cite{ABCPV96} this method suffers of the problem that 
when $\delta_n$ is finite one is not sure to sample correctly the measure on the 
statistically stationary state. Indeed a finite perturbation will generically bring 
the trajectory out from the ``attractor''. Nevertheless the result is in good agreement 
with the one obtained in \cite{Kantz} and confirms that at the origin 
of the perturbation propagation in this system there should be a mechanism very 
similar to the one discussed for the shift map. Further studies related to the
``stable chaos'' phenomenon have been recently performed~\cite{ginelli}.

\section{Analogies with front propagation in reaction diffusions systems}
\label{sec:5}
The perturbation evolution in spatially distributed systems 
can be described as the motion of an interface separating 
perturbed from unperturbed regions. 
In this spirit, one can wonder if and to what extent it is possible to draw
an analogy between the evolution of this kind of interface and the propagation 
of fronts connecting steady states in reaction diffusion systems. 
As already noticed in Ref.~\cite{TGP95}, the two phenomena display many similarities.
In the following we will discuss the similarities and differences, in particular 
we will introduce a simple phenomenological model which can help us in highlighting
the analogies.

Let us start by recalling the basic features of fronts propagation in 
reaction diffusion systems  with reference to
the Fisher-Kolmogorov-Petrovsky-Piscounov (FKPP) equation \cite{KPP}
\begin{equation}
\partial_t \theta(z,t) = D\partial^2_{zz}\theta(z,t) + G(\theta(z,t))\,,
\label{eq:kpp}
\end{equation}
where $\theta(z,t)$ represents the concentration of a reactants which diffuses and 
reacts, the chemical kinetics is given by $G(\theta)$. Typically the function 
$G(\theta)\in C^1[0,1]$ (with $G(0)=G(1)=0$) exhibits
one stable ($\theta=1$) and one unstable ($\theta=0$) fixed point.
Once the system is prepared on the stable state ($\theta(z)=0 \, \forall z$),
an initial (sufficiently steep) perturbation (e.g. a step function
$\theta(z,t=0) = \Theta{(z-z_0)}$ ) will give rise to a smooth front 
moving with a velocity $V_p$, that will connect
the unstable and the stable fixed points:
as a result the stable state will invade the unstable one.
This equation admits many different traveling solutions that
are typically characterized by their propagation velocities,
however for a sufficiently steep initial perturbation
of the unstable state the selected front is unique
and its velocity $V_p$ is bounded in the interval $[V_{min}, V_{max}]$,
where
\begin{equation}
V_{min}=2\sqrt{D G^{'}(0)}
\label{eq:vmin}
\end{equation}
and
\begin{equation}
V_{max}=2\sqrt{D
\sup_{0<\theta<1}\left\{ {G(\theta) \over \theta} \right\}}\,.
\label{eq:vmax}
\end{equation}  
If $G(\theta)$ is concave $\sup_{0<\theta<1}\{G(\theta)/\theta\}$ is attained
at $\theta=0$ (i.e. $\sup_{0<\theta<1}\{G(\theta)/\theta\}=G^{'}(0)$) and the
selected velocity is always the minimum one: the front is ``pulled'' by the 
growth of infinitesimal perturbations of the unstable state.
Otherwise, if $G(\theta)$ is convex one can observe a velocity
$V_p > V_{min}$: the front is now ``pushed'' by the growth 
of finite amplitude perturbations. These
results are known as the Aronson \& Weinberger theorem
\cite{ara}
(for more details see Ref. \cite{wim2}).
The velocity $V_{min}$ can be easily obtained by performing a
linear analysis of (\ref{eq:kpp}) and by employing a marginal
stability criterion \cite{wim1}.

The subject of our analysis is the spreading of perturbations 
in a chaotic media. In order to compare our case with
the FKPP, one we should consider the time evolution of the difference 
of two chaotic trajectories $\{\Delta x_i(t)\}_{i=1,L}$. 
However, the nature of the two phases separated by the front is now different from 
the FKPP case. 
The interface separates an unstable ($\Delta x_i=0$) state from a ``statistically
stable'' one. With the term ``statistically stable'' we mean that behind the front,
in the bulk of the perturbed region, $\Delta x_i(t)$ does not converge to a stable 
fixed point (as for the FKPP) but it fluctuates in a stationary way
around an average value. This suggests that a good model for reproducing
this dynamical evolution would be a FKPP with a stochastic kinetics~\cite{armero}.
However, many similarities with FKPP can be
established by neglecting the chaotic fluctuations and
considering the average shape of the front
\cite{TGP95}. In practice, this can be done by averaging
the perturbation evolution over many different initial conditions.

Once the chaotic fluctuations are neglected, one can 
express the average perturbation growth in any site of the 
chain via the following mean-field approximation:
\begin{equation}
u_i(t+1)=\exp[\lambda(\tilde{u}_i(t))]\, \tilde{u}_i(t)\,,
\label{eq:model}
\end{equation}
where $i$, $t$ are the discrete space and time indices and 
$\tilde{u}_i=(1-\varepsilon)u_i+\varepsilon/2(u_{i+1}+u_{i-1})$. Here, 
$u_i(t)$ indicates the ``average'' 
$\Delta x_i(t)$ and  $\lambda(u)$ the corresponding
FSLE (at least for positive growth rates).
In the limit $u\to 0$  $\lambda(u)\to \lambda$, while,
for $u \sim {\cal O}(1)$, $\lambda(u)$ should reflect the 
saturation effects associated to the nonlinear map. 
In the infinitesimal limit ($u \to 0$)
one essentially recovers the model discussed in Ref.~\cite{PP98}.

By passing to continuous variables is possible to
show that Eq.~(\ref{eq:model}) can be 
reduced to Eq.~(\ref{eq:kpp}) (at least at the
leading order).
In order to transform Eq.~(\ref{eq:model}) in its
continuous version,
let us introduce infinitesimal spatial and temporal 
resolutions $dx$ and $dt$, and  assume that the
diffusive scaling holds, i.e. $dx^2=dt$. 
Limiting to the first order expansion in $dt$
(second order in $dx$), one easily shows that
the continuous counterpart of Eq.~(\ref{eq:model}) is
\begin{equation}
\partial_t u= \lambda(u)u+\frac{\varepsilon}{2} \partial^2_{xx}u\,,
\label{eq:model_kpp} 
\end{equation}
which is nothing but (\ref{eq:kpp}) with 
\begin{equation}
\lambda(u)={G(u)\over u}\,,
\label{eq:mapping}
\end{equation}
and $D=\varepsilon/2$. 
As a consequence of the identity (\ref{eq:mapping}),
one should observe a pulled dynamics for the
chaotic front if $\lambda \ge \lambda(u) \quad \forall u \ge 0$,
and a pushed one could occur
only if $\max_{u}\{\lambda(u)\} > \lambda$.
This can be considered as a reformulation of the 
Aronson \& Weinberger theorem \cite{ara} in the
context of information propagation.
The velocity bounds (\ref{eq:vmin}),
(\ref{eq:vmax}) can be now identified with 
$V_a = \sqrt{2\varepsilon \lambda}$, i.e. Eq.~(\ref{eq:velocity}), 
the first one and with
\begin{equation}
V_s=\sqrt{2\varepsilon \max_{u}\{\lambda(u)\}}
\end{equation}
the second one. We stress again that the lower velocity
bound is indeed represented by $V_L$ and that it
coincides with $V_a$ only for sufficiently strong 
diffusive coupling $\varepsilon$.

Another interesting point that we can investigate via a 
numerical simulation of the mean field effective 
equation (\ref{eq:model})
is the dependence of the propagation properties on the
specific shape of $\lambda(u)$. 
For instance, as previously conjectured, we expect that 
if a monotonous decreasing $\lambda(u)$ is considered
one should observe linear propagation, only.
Indeed, numerical integrations of the model (\ref{eq:model}), 
with the choice $\lambda(u)=A-Bu$, being $A$ and $B$ positive
constants, show that $V_p =V_a= \sqrt{2 \varepsilon A}$,
where $A$ corresponds to the maximal Lyapunov of the effective model.
A generalization of this simple model would require to 
consider $A$ and $B$ as fluctuating quantities generated by
suitable stochastic processes. But while for the choice 
of $A$ one can have some hint \cite{PP98}, this is not 
the case for $B$. 

\begin{figure}
\epsfxsize=8.truecm
\epsfysize=6truecm
\epsfbox{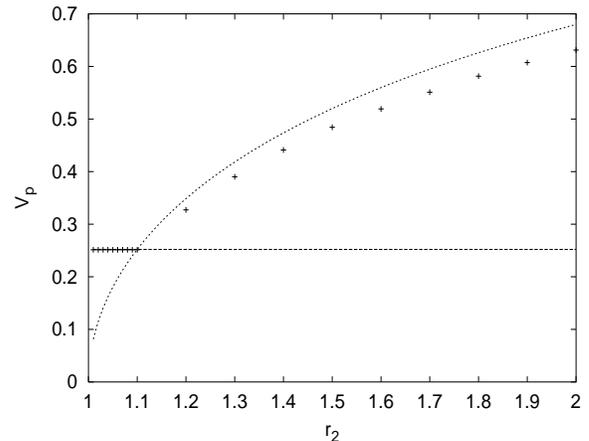}
\caption{Propagation velocities 
(symbols) for the model (\ref{eq:model}) 
with $\lambda(u)$ given by
(\ref{eq:mod1}) as a function of $B = \ln(r_2)$,
with $\varepsilon=1/3$,  $A=\ln{1.1}$ and
$C=A$. The threshold  values are fixed to
$\delta^{NL}= 10^{-3}$ and $\delta^{sat}=0.53$
and a chain of $4 \cdot 10^4$ sites has been used.
The curves refer to $V_{a}=\sqrt{2\varepsilon A}$ (solid line)
and $V_{s}=\sqrt{2\varepsilon B}$ (dashed line).
Note that in the linear case, $B<A$,
there is a perfect agreement between the measured velocity and the
prediction. 
}
\label{fig:vlvnl_model}  
\end{figure}
Let us now consider the non linear propagation case, as we have
previously seen a necessary condition in order
to have $ V_p > V_L$ is that $\max_{u}\{\lambda(u)\} > \lambda$
for some finite $u$. For what concerns the actual value of the
selected velocity, this will depend on the value of the diffusive
coupling and on the specific shape of $\lambda(u)$. 
In the following we will examine the dependence of 
$V_p$ on two quantities that characterize $\lambda(u)$:
the difference $|\max_{\delta} \lambda(\delta) - \lambda|$ and the
scale $\delta^{NL}$ at which the non linear effects set in.

As a first example let us consider $\lambda(u)$ as a step function:
\begin{equation}
\lambda(u)=\left\{ \begin{array}{ll}
A & 0<u<\delta^{NL}\\ 
B & \delta^{NL}\leq u <\delta^{sat}\\ 
-C & u\geq \delta^{sat}\,,
\end{array}
\right.
\label{eq:mod1}
\end{equation}
where $A$, $B$ and $C$ are positive quantities, with $B > A$,
while $\delta^{NL}$ and $\delta^{sat}$ are amplitude
thresholds. The parameter $A$ is nothing but the Lyapunov exponent, 
$B$ mimics the non linear terms leading to
an enhancement of the growth rate, and the last term mimics the 
damping of the perturbation due to the saturation effects.
In Fig.~\ref{fig:vlvnl_model} it is reported 
the behavior of the propagation velocity
for the model (\ref{eq:model}) with $\lambda(u)$ given by
(\ref{eq:mod1}) for various values of $B$, once $A$ and $C$ are
fixed. From the figure is evident that if $B < A$ then 
$V_p \equiv V_a$ (i.e. we are in the linear regime), 
while as soon as $B > A$ an increase of $V_p$
with respect to $V_a$ is observed.

\begin{figure}
\epsfxsize=8.truecm
\epsfysize=6truecm
\epsfbox{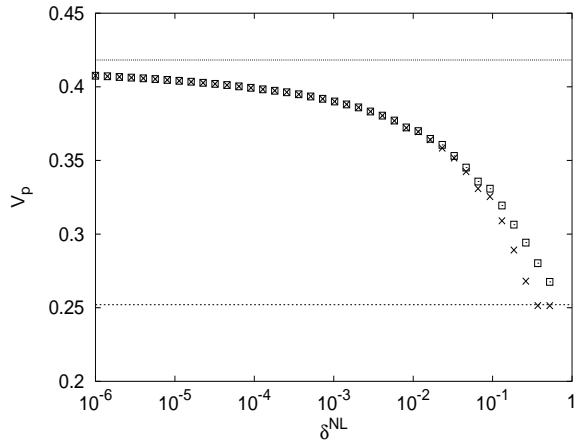}
\caption{Propagation velocities $V_p$ as a function of $\delta^{NL}$
for the model (\ref{eq:model}), (\ref{eq:mod1}) with $\varepsilon=1/3$, 
$A=\ln(1.1)$, $B=\ln(1.3)$, $C=A$, $\delta^{sat}_0=0.53$. 
The crosses refer to the case $\delta^{sat}=\delta^{sat}_0$,
while the boxes to the case $\delta^{sat}=\delta_0^{sat}+\delta^{NL}$.
The two lines correspond to $V_{a}=\sqrt{2\varepsilon A}$ and 
$V_{s}=\sqrt{2\varepsilon B}$.}
\label{fig:step}  
\end{figure}
 In the whole examined parameter 
range $V_p$ is always reasonably approximated by $V_s$, but smaller.
By increasing $B$ one observes an increase of the difference
between the measured velocity and the linear prediction.
These results confirm that the condition (\ref{eq:nonlin}) is indeed a
necessary condition in order to observe nonlinear propagation
of information. 

We will now investigate the role of $\delta^{NL}$ 
in determining the propagation velocity. It is quite obvious that
modeling the dynamics via the Eqs.~(\ref{eq:model}) and
(\ref{eq:mod1}) the nonlinear effects will disappear in
the limit $\delta_{NL}\to \delta^{sat}$ and this is indeed
confirmed by the simulations (see Fig. \ref{fig:step}).
In order to examine a less obvious situation,
a modification of the expression (\ref{eq:mod1})
is also considered. In this second case 
$\delta^{sat}=\delta^{sat}_0+\delta^{NL}$
is assumed, therefore by varying $\delta^{NL}$
the extension of the amplitude interval over which the nonlinear
mechanism is active will not be modified (being fixed to $\delta^{sat}_0$).
But also in this second case $V_p \to V_a$ for increasing
$\delta^{NL}$, this indicates that the smaller are the perturbation
amplitudes affected by the nonlinear mechanisms
the stronger will be the nonlinear effect on the velocity :
$V_p \to V_s$ for $\delta^{NL} \to 0$. 

Typically, in generic CML's by varying a control parameter 
both the difference $|\max_{\delta} \lambda(\delta) - \lambda|$  
as well as $\delta^{NL}$ will change. Moreover,
even the definition of $\delta^{NL}$ for a continuous 
$\lambda(\delta)$ is not obvious. In order to understand
the validity of the mean-field approximation
(\ref{eq:mod1}) in a more realistic case we consider
the shift map. In particular, for $\lambda(u)$ we
employed the analytical expression (\ref{eq:lambda_shift})
valid for the single uncoupled map.

\begin{figure}
\epsfxsize=8.truecm
\epsfysize=6truecm
\epsfbox{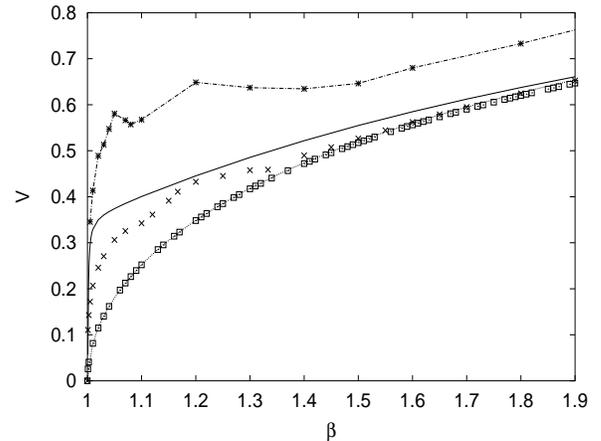}
\caption{Information propagation velocities for the shift map with 
$\varepsilon=1/3$: (boxes) linear velocities $V_L$ and (crosses) directly
measured nonlinear ones. The two lines correspond to
$V_{a}$ (\ref{eq:velocity}) (dotted line) 
and to the propagation velocity for the model 
(\ref{eq:model}) with $\lambda(u)$ 
given by (\ref{eq:lambda_shift}) (solid line). The dashed curve with asterisks is 
 $V_{s}=\sqrt{2 \varepsilon \max_{\delta}\{\lambda(\delta)\}}$. }
\label{fig:vlvnl_shift}  
\end{figure}

The FSLE for coupled shift maps is actually different from 
(\ref{eq:lambda_shift}), but we are neglecting correlations 
among different sites and effects due to the specific measure 
associated to the system. 
Nevertheless, a numerical integration of Eq.~(\ref{eq:model})
equipped with (\ref{eq:lambda_shift}) reproduces 
semi-quantitatively the 
features observed for a lattice of coupled shift maps 
(see Fig.~\ref{fig:vlvnl_shift}). In particular, also the
simple mean-field model (\ref{eq:model}) with the choice
(\ref{eq:lambda_shift}) is able to forecast the
observed transition
from nonlinear to linear behavior for $\beta \to 2$.

Concluding this section we can safely affirm that (\ref{eq:model}) is a 
reasonable 
model to mimic the perturbation evolution at a mean field level, neglecting
the spatio-temporal fluctuations and correlations. In the same fashion
$\lambda(u)$ can be considered as an ``effective'' non linear kinetics 
for the perturbation evolution.

\section{Final remarks}
\label{sec:6}

In this paper information (error) propagation in extended chaotic
systems has been studied in details. In particular, we have analyzed
the relevance of linear and nonlinear mechanisms 
for the propagation phenomena in spatio-temporal chaotic
coupled map lattices. Linear stability analysis is
not always able to fully characterize disturbance propagation.
This is particularly true for (marginally) stable systems
with strong nonlinearities,
where finite size perturbations are responsible for information
spreading in the system. When the nonlinear effects prevail
on the linear ones the propagation velocity of information
$V_p$ can be higher than the linear velocity $V_L$. A
necessary condition for the occurrence of information
spreading induced by nonlinear mechanisms has been
expressed in terms of the Finite Size Lyapunov Exponent (FSLE).
We have also shown the existence of strong analogies between
error propagation and front propagation in reaction-diffusion
models. In particular, the above mentioned necessary condition
is analogous to the Aronson \& Weinberger theorem \cite{ara} for 
front connecting stable and unstable steady states.
In the linear and nonlinear case, the propagation velocity 
$V_p$ can be identified via an unique marginal stability criterion
involving Finite Size Lyapunov Exponents defined in a
moving reference frame. This result generalizes the 
corresponding linear criterion expressed in terms of 
the maximal comoving Lyapunov exponents
for the identification of $V_L$ \cite{DK87}. 

These results can be of some interest for
the synchronization and the control of extended 
systems. It has been recently shown that
the synchronization of coupled extended 
systems is strongly influenced
by nonlinear effects. In particular, 
the synchronization time exponentially
diverges with the system size, even in non-chaotic
situations, provided that $V_p > 0$ \cite{noi}.
We believe that in these systems the appropriate
indicator to characterize such transition
would be the ``transverse'' Lyapunov 
exponent~\cite{pecora}, once extended to finite scales.
As far as control schemes are concerned, since
they rely mainly on linear analysis~\cite{ogy},
new nonlinear methods (e.g. based on the concept of FSLEs)
should be introduced in order to control the erratic 
behaviors due to fully nonlinear mechanisms.

A further aspect that should be addressed in future work
concerns the extension of the applicability of the FSLE
also to linearly stable systems: a candidate in this
respect could be the indicator recently introduced
in \cite{Kantz}. 

Finally, it is reasonable to expect that the
present analysis is not limited to discrete
models but that it can be applied to continuous 
extended systems described in terms of PDE's,
e.g. to the complex Ginzburg-Landau equation.
\acknowledgments
Stimulating interactions with F.~Cecconi, R.~Livi, K.~Kaneko,
A.~Pikovsky, A.~Politi, and A.~Vulpiani are gratefully acknowledged.
Part of this work has been developed at the Institute of Scientific 
Interchange in Torino, during the workshop on ``Complexity and
Chaos'' in October 1999.
We acknowledge CINECA in Bologna and INFM for providing us access to the 
parallel CRAY T3E computer under the grant ``Iniziativa Calcolo
Parallelo''.

\begin{appendix}
\section{}

In any numerical computation of the Lyapunov exponents one is forced to
use a finite time approximation for an infinite time limit. 
Nevertheless, provided that the convergence to the asymptotic value 
is fast enough, this is not a dramatic problem.
As a matter of fact, in low dimensional systems
very fast convergence to the asymptotic value is
usually observed.
However, this problem manifest more
dramatically in high dimensional systems, where the time to align along 
the direction of maximal expansion could be very long \cite{PP98,P93,Orszag}.

In the present case this problem is
complicated by the fact that the FSLE is intrinsically a finite time 
indicator. Indeed the time a perturbation takes to grow from a
value $\delta$ to $r \delta$ is finite unless $\delta \to 0$.
However, for the CML models here analyzed and for initially
localized perturbation (\ref{eq:tzero}) it is possible to evaluate 
the corrections to apply to the FSCLE, estimated at
finite time, in order to recover, for sufficiently small $\delta$,  
the expected limit $\Lambda(v)$. 

These corrections allow for a faster convergence
of the FSCLE to its asymptotic values.

For the sake of simplicity we consider maps with constant slope, 
e.g. the shift map $F(x)=\beta\, x\;\;{\mbox {mod}}\,1$,
and the CML defined in Eq.~(\ref{eq:cml}).
In this case (for $\varepsilon < 1/2$) the maximal Lyapunov exponent
of the CML coincides with that of the single map \cite{isola};
i.e.  $\lambda=\ln \beta$. We will limit to the case
$v=0$ (i.e to the FSLE), since the extension to generic $v$ is 
straightforward.
As shown in ~\cite{P93,lyapcorr} the finite time evolution
of an infinitesimal perturbation $d_0$ initially
localized in $i=0$ can be expressed at time $T$ as the sum of the 
contributions $M(m,T)$ associated to all the paths connecting
the space-time point $(i=0,t=0)$  to the point $(i=0,t=T)$,
i.e.
\begin{equation}
\frac{\delta x_{i=0}(T)}{d_0} = \beta^T \sum_{m=0}^T M(m,T)  \quad .
\label{path}
\end{equation}
where $m$ is the number of ``diagonal links'' connecting
$(i,t)$ to $(i \pm 1,t+1)$ present in the path of length $T$.
Each path $[m,T]$ of length $T$ with $m$ diagonal links contributes
to the above sum with a term
\begin{equation}
M(m,T) = N(m,T)
\left(\frac{\varepsilon}{2}\right)^m (1-\varepsilon)^{T-m}
\end{equation}
where $N(m,T)$ is the multiplicity associated to each path 
$[m,T]$ (for more details see \cite{lyapcorr}).
As can see from Eq.~(\ref{path}) the {\it finite time Lyapunov
exponent} $\lambda_T$ will be given by
\begin{equation}
\lambda_T=
\frac{1}{T}\ln\left|\frac{\delta x_{0}(T)}{d_0}\right| = \ln \beta + 
\frac{1}{T}\ln\left|\sum_{m=0}^T M(m,T)\right|
\quad .
\label{lyt}
\end{equation}
the first contribution is the asymptotic one, while 
the second one will vanish in the limit $T \to \infty$
and it is the finite-time correction to evaluate.
This second contribution can be numerically estimated
by considering the finite time evolution of the Lyapunov
eigenvector $\{W_i(t)\}$ associated to the maximal Lyapunov exponent,
once it is initialized as $W_i(0) = d_0 \, \delta_{i,0}$.
The evolution of $\{W_i(t)\}$ in the tangent space 
is ruled by the following equation 
\begin{equation}
W_i(t+1)=\beta \left[(1-\varepsilon) W_i(t)+
{\varepsilon \over 2} (W_{i-1}(t)+W_{i+1}(t))\right]\,.
\label{eq:ac.tangent}
\end{equation}
In order to evaluate the finite time corrections,
one should iterate at the same time 
Eq.~(\ref{eq:ac.tangent}) (with $\beta$
fixed to one) and the two replicas 
required for computing $\lambda(\delta)$ (see Sect.~\ref{sec:2}).
Then the estimation of the FSLE should modified in the following way:
\begin{eqnarray} 
\lambda(\delta)={1 \over \langle \tau(\delta_n,r)\rangle_e }
 \left\langle \ln \left( { |\delta x_i(t+\tau(\delta_n,r))| \over |\delta
 x_i(t)| }  \right) \right. \nonumber \\
\left.  - 
\ln \left( { |W_i(t+\tau(\delta_n,r))| \over |W_i(t)|}
\right) \right\rangle_e \,.  
\label{eq:FSLE_discrete_new}
\end{eqnarray}
The case of maps with non constant slope is computationally much
heavier. Since for each different path the local multiplier
$F^{'}(\tilde{x_i})$ will be different and they will depend
on the particular trajectory under consideration \cite{P93,lyapcorr}.
As a matter of fact we have observed that if the multipliers are
equally distributed among positive and negative values, the finite
time corrections essentially cancel out.
\end{appendix}


\end{multicols}


\begin{thebibliography}{99}
\bibitem{nota}Note that also for $\varepsilon \to 1$ deviations are observed. 
This is due to the fact that for $\varepsilon=1$ the system reduces to
two uncoupled subchains, therefore also for $\varepsilon \sim 1$
the coupling among nearest neighbors is actually very small.
\end{thebibliography}
\end{document}